\documentclass[aps, pre, twocolumn,superscriptaddress,amsmath,amssymb,floatfix]{revtex4-1}

\usepackage{graphicx}
\usepackage[usenames]{color} 
\usepackage{comment} 
\usepackage{bm}
\usepackage{amsmath}
\newcounter{abc}

\newcommand{\be}{\begin{equation}} 
\newcommand{\ee}{\end{equation}}
\newcommand{\bea}{\begin{eqnarray}} 
\newcommand{\eea}{\end{eqnarray}}
\newcommand{\nav}{\bar{n}_s}
\newcommand{\navsq}{\overline{n_s^2}}
\newcommand{\D}{\Delta}
\newcommand{\emin}{e_{\rm min}}

\begin{document}

\title{Demixing cascades in cluster crystals}

\author{Nigel B. Wilding} 
\affiliation{Department of Physics, University of Bath, Bath BA2 7AY,
United Kingdom} 
\author{Peter Sollich}  
\affiliation{King's College London, Department of Mathematics, Strand,  London WC2R 2LS,
United Kingdom} 

\begin{abstract}

In a cluster crystal, each lattice site is occupied by multiple
soft-core particles. As the number density is increased at zero
temperature, a `cascade' of isostructural phase transitions can occur
between states whose site occupancy differs by unity. For low but
finite temperature, each of these transitions terminates in a critical
point. Using tailored Monte Carlo simulation techniques we have
studied such demixing cascades in systems of soft particles
interacting via potentials of the generalized exponential form
$u(r)=\epsilon\exp[-(r/\sigma)^n]$.  We have estimated the critical
parameters of the first few transitions in the cascade as a function
of the softness parameter $n$. The critical temperature and pressure
exhibit non-monotonic behaviour as $n$ is varied, although the
critical chemical potential remains monotonic. The trends for the
pressure and chemical potential are confirmed by cell model
calculations at zero temperature. As $n\to 2^+$, all the transitions
that we have observed are preempted by melting although we cannot rule
out that clustering transitions survive at high density.

\end{abstract}

\maketitle

\section{Introduction}

Soft matter systems such as star polymers and dendrimers comprise
individual molecules that can overlap substantially at high
concentrations \cite{Likos:2006ys,Likos:2007fk}.  In order to better
understand the equilibrium and dynamical properties of such systems
one generally appeals to theory and simulation. In so doing it is common
to dispense with the finer (atomistic) detail in favour of
coarse-grained descriptions.  Typically these represent each molecule
in terms of an ultra-soft colloidal particle which interacts with its
neighbours via a short ranged two-body effective potential.  
The form of this potential can be parameterized from simulation and
experiment. For instance, for star polymers in good solvent one finds
a weakly divergent repulsive potential \cite{Jusufi:1999ve}. However,
if the monomer density is sufficiently low that the centres of mass
can coincide, a bounded potential is appropriate \cite{Likos:2001fk}.

\begin{figure}[h]
\includegraphics[type=pdf,ext=.pdf,read=.pdf,width=0.95\columnwidth,clip=true]{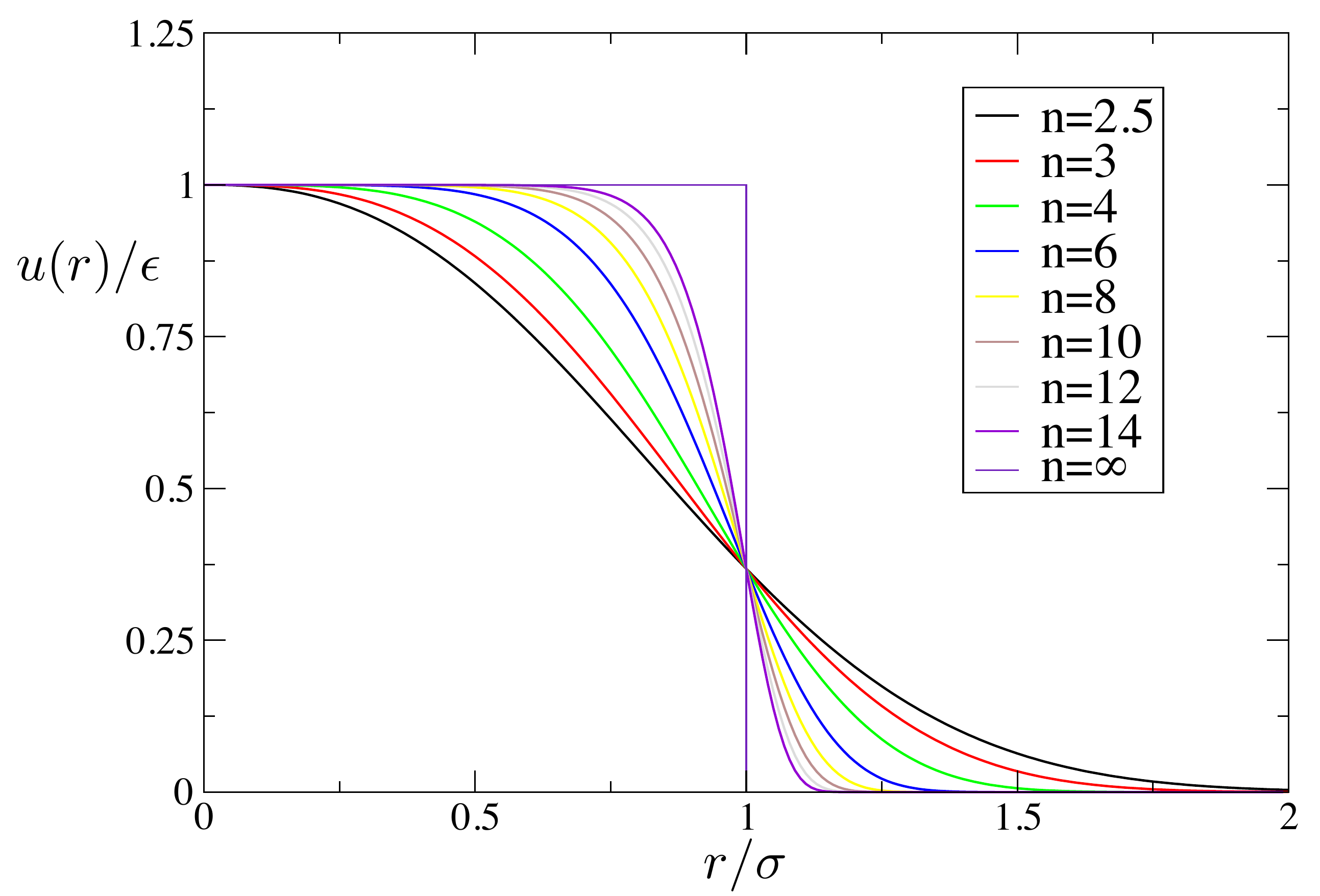}
\caption{The GEM-$n$ potentials $u(r)=\epsilon\exp[-(r/\sigma)^n]$, shown for the values of the softness parameter $n$ studied in this work.}
\label{fig:GEM-n}
\end{figure}

Systems described by bounded interactions have received considerable
attention in recent years due to their unique equilibrium and
dynamical behaviour. A prototype theoretical form for a bounded
potential is the generalized exponential model (GEM) for which the
interaction potential is given by
\be
u(r)=\epsilon\exp[-(r/\sigma)^n]\:.
\ee
Here $\epsilon$ and $\sigma$ set the energy and length scales
respectively, while $n$ is a `softness' parameter which also serves to
delineate the members of the GEM-$n$ class of models. Certain  members of this class have
been extensively investigated by several groups
\cite{Mladek:2006ys,Mladek:2008fk,Zhang:2010fk,Zhang:2012ek,Suto:2011vn,Coslovich:2013uq}.
For $n=2$ the potential is a simple Gaussian and the model is termed the
Gaussian core model (GCM)
\cite{Stillinger:1976fv,Prestipino:2005uq,Lang:2000ij,Likos:2001uq,Ikeda:2011qy,Coslovich:2013uq};
while for $n=\infty$ one obtains a top hat potential known as the
penetrable sphere model (PSM)
\cite{Blum:1979lq,Marquest:1989ul,Likos:1998rw}. Various members of
the spectrum of GEM$-n$ potentials are depicted in
Fig.~\ref{fig:GEM-n}.

The key feature of the equilibrium behaviour of particles interacting
via the GEM-$n$ potential is that for $n>2$ they exhibit {\em
  clustering} behaviour in which particles clump together in
groups. This phenomenon (the origin of which can be traced to
instabilities associated with negative components in the Fourier
transform of the pair potential \cite{Likos:2001uq}) is already
evident in dense liquids \cite{klein:1994fk,Fragner2007}, but is most
striking in the crystalline phases where lattice sites are occupied by
multiple particles \cite{Likos:1998rw,Mladek:2006ys,Mladek:2008fk}.
Activated hopping
\cite{Moreno:2007zm,Montes-Saralegui:2013ec} of particles between
lattice sites contributes to density fluctuations and dynamical relaxation processes in such systems. Although originally only observed in the GEM$-n$
models, evidence for cluster crystals has recently been reported in
simulations of dendrimer models with atomistic detail
\cite{Lenz:2012fk}. To date, however, there have been no experimental
reports of cluster crystals in real soft matter systems.

Most studies of the GEM-$n$ family have been performed for the three
cases $n=2,4,\infty$. In the GCM ($n=2$)
\cite{Stillinger:1976fv,Prestipino:2005uq,Lang:2000ij,Likos:2001uq,Ikeda:2011qy,Coslovich:2013uq},
no clustering occurs, but the system exhibits two solid phases, one
face centred cubic (fcc) and the other body centred cubic (bcc). As
the density is increased at low temperature, reentrant melting occurs
so that the highest density state is always a fluid.

For $n=\infty$ (the PSM), clustering is observed in both the fluid and
solid phases
\cite{Likos:1998rw,Schmidt:1999tw,Suh:2010hq,Fernaud:2000zt,zhangkai:2012gb}. Although
only one phase transition has been reported to date, namely the
liquid-solid transition, the freezing properties are rather
interesting because they exhibit crossover behaviour depending on
temperature. Specifically, at high $T$ clusters form in the liquid and
these freeze into a cluster crystal. By contrast at sufficiently low
temperature the interparticle potential reduces to that of a system of
hard spheres and the freezing transition behaves accordingly.

The GEM-$4$ potential is the most studied member of the GEM-$n$ family
to date, see
eg.\ \cite{Likos:2007fk,Mladek:2006ys,Mladek:2008fk,Neuhaus:2011fk,Zhang:2012ek}. This
system exhibits a rich phase diagram including bcc and fcc cluster
phases, as well as reentrant phase behavior. At low temperature,
evidence has been found for an infinite cascade of isostructural
demixing transitions between fcc phases having different site
occupancies.  \cite{Neuhaus:2011fk,Wilding:2013qv}.

Previously we located the critical points of the first four
stable transitions in the cascade for the GEM-4 potential, ie.\ those which
at $T=0$ exhibit unit jumps in the site occupancy
$n_s=2\leftrightarrow 3$, $n_s=3\leftrightarrow 4$,
$n_s=4\leftrightarrow 5$, $n_s=5\leftrightarrow 6$. Interestingly,
within simulation uncertainties, no variation was seen in the critical
temperatures for these four transitions. In the present work we extend our
investigations to other members of the GEM-$n$ class. Our aim is to
determine whether the demixing cascade seen for $n=4$ persists for
other values of $n$ and, if so, how the critical parameters depend on
$n$.  Additionally we seek to understand the fate of the demixing
transitions as one approaches the Gaussian limit ($n=2$), for which no
cluster crystals appear to exist.  We also consider the case of
large $n$ in which the potential approaches the PSM limit.

\section{Methods}

\label{sec:methods}
\subsection{Monte Carlo Scheme}
\label{sec:MC}

Crystals in which the number of particles per lattice site can vary,
are not straightforward to study by simulation. To appreciate why,
consider a system comprising $N$ particles in a volume $V$. Suppose
there are $N_s$ lattice sites so that the average occupancy is
$n_s=N/N_s$ and the volume per lattice site is $v_s=V/N_s$. Then the
particle number density is simply

\be 
\rho=\frac{N}{V}=\frac{n_s}{v_s} \:.
\ee 

Clearly, however, a given $\rho$ can be realized by an infinite number
of combinations of $n_s$ and $v_s$. Equilibrium corresponds to
each lattice site having a certain occupancy $n_s^{\rm eq}$ and a certain
unit cell volume $v_s^{\rm eq}$. But in order to relax to this state from
some arbitrary initial state, it is in general necessary for the number of lattice sites,
$N_s$, and the lattice parameter $a$ to change.

Unfortunately, fluctuations in $N_s$ do not typically occur on
simulation timescales. For a system having periodic boundary
conditions, $N_s$ can vary only if a whole crystal plane is added or
deleted.  But free energy barriers prevent such large changes from
happening. This is true even if one operates in an ensemble in which
the system volume (and hence the lattice parameter) can
fluctuate. Accordingly, if the system is initiated with a given number
of lattice sites, it generally remains so for the duration of the
simulation. Even if plane insertions/deletions were to occur, for a
finite-sized system the consequent large relative changes in $N_s$
would lead to considerable discretisation effects in the values of
$n_s$ which could be sampled.

In order to locate the equilibrium conditions, a different strategy
must be taken. Specifically it has been shown
\cite{Swope:1992fk,Mladek2007} that equilibrium corresponds to the condition

\be
\mu_s=0\:,
\ee
where $\mu_s$ is the so-called lattice site (or cluster) chemical potential given by
\be
N_s\mu_s=F+PV-\mu N \:,
\label{eq:cluschempot}
\ee
with $F$ is the Helmholtz free energy, $P$ is the pressure and
and $\mu$ is the standard chemical potential. 

Unfortunately $\mu_s$ cannot be directly measured as a simple ensemble
average at the state point of interest and therefore one must resort
to more elaborate means. One approach for estimating $\mu_s$ is a
direct assault on the right hand side of Eq.~(\ref{eq:cluschempot})
\cite{Mladek2007}: obtaining $F$ via thermodynamic integration from a
reference state of known free energy, $P$ by sampling the virial and $\mu$
using the Widom insertion method~\cite{FrenkelSmit2002}. This process
(or alternatively a direct estimation of the constrained free energy
\cite{Zhang:2010fk}), then has to be repeated for a range of values of
$n_{s}$ in order to pinpoint equilibrium at the prescribed
$\rho$. Accordingly it can be cumbersome and laborious.

In recent work we have proposed a new Monte Carlo simulation scheme
for efficiently and accurately locating the equilibrium conditions in
cluster crystals. The method is framed within the great grand
canonical (constant $\mu,P,T$) ensemble. For solids having fixed $N_s$
(a constraint imposed implicitly by free energy barriers, as described
above), this ensemble does not suffer from the divergence of the partition
function that occurs in equilibrium fluids \cite{Wilding:2013qv}.  One
benefit of its use is that it is fully unconstrained, allowing
fluctuations in $N,V,E$: fluctuations in $V$ permit the relaxation of
the lattice parameter, while fluctuations in $N$ allow the average
site occupation $\nav=N/N_s$ to vary in small steps of $1/N_s$.
Another advantage is that the great grand canonical ensemble permits the ready use
of histogram reweighting to scan the fields $\mu,P,T$, without the
need for multiple simulations.

In order to locate equilibrium we implement a MC move that permits fluctuations in the
number of lattice sites. Specifically, we define two states of the
system, $\alpha=0$ and $\alpha=1$, which differ by a single lattice
plane of $M_s$ lattice sites. For $\alpha=0$ the number of lattice
sites is $N_s^{(0)}=N_s+M_s$, while for $\alpha=1$ it is
$N_s^{(1)}=N_s$. Biased sampling techniques are used to access regions
of configuration space that allow a lattice plane to be 'switched' in
and out of the system via a Monte Carlo update. This back and forth
switching between the $\alpha=0$ and $\alpha=1$ states allows one to
measure the relative probability of finding the system in the
$\alpha=0$ and $\alpha=1$ states

\be
{\cal R}=\frac{p^{(1)}}{p^{(0)}}\:.
\ee
It can be shown \cite{Wilding:2013qv} that this probability ratio provides direct access to the difference

\be
\ln({\cal R})=(N_s^{(1)}-N_s^{(0)})\mu_s\:,
\ee
and since the right hand side vanishes only when $\mu_s=0$, this
allows the equilibrium conditions to be estimated via an equal peak
weight criterion: ${\cal R}=1$. In practice one locates equilibrium with the help of histogram
reweighting, varying $\mu$ and $P$ together at fixed $T$ in such a way
as to maintain some target density. The equal peak weight criterion
identifies the specific combination of $\mu$ and $P$ that corresponds to equilibrium at this density. For further details the interested reader is referred to
Ref.~\cite{Wilding:2013qv}.

\subsection{Cell model}
\label{sec:cellmodel}

We can study the zero temperature behaviour of demixing cascades in
the GEM-$n$ models using a simple cell model inspired by
Refs.~\cite{Likos:1998rw,Neuhaus:2011fk}. We assume that the crystal consists of
$N_s$ sites as above, and is substitutionally disordered in the sense
that the number of particles $n_s$ at each site is drawn from some
distribution $p_{n_s}$. If we also assume that at $T=0$ the particles
sit at the lattice positions, then in units where $\epsilon=\sigma=1$
we can write down the energy of such a crystal as

\begin{eqnarray}
E&=& \sum_{n_s} p_{n_s} N_s \frac{1}{2}n_s(n_s-1)\nonumber\\
&\:&
+\frac{1}{2} \sum_{n_s} p_{n_s} N_s n_s z \nav u(d)\:.
\end{eqnarray}

The two terms describe interactions between particles on the same and
on different sites, respectively.  The distance between neighbouring
lattice sites is $d=a/\sqrt{2}$ in an fcc lattice, or $d=ca$ more
generally, where $a$ is the lattice parameter. We have also denoted by
$z$ the coordination number of the crystal lattice, and by $\nav =
\sum_{n_s} p_{n_s} n_s$ the average number of particles per site. To
express $E$ in terms of $\rho$ and the distribution of cluster sizes $n_s$, one uses
$N/N_s = \nav$ and $N=\rho L^3$, with $L$ the linear system size. A
third relation is $N_s = A(L/a)^3$ where $A$ indicates the number of
particles per cubic unit cell, with $A=4$ for fcc. This gives
$a=(A\nav/\rho)^{1/3}$ and overall for the energy density

\begin{equation}
e=\frac{E}{L^3} = \frac{\rho}{2}\left[\frac{\navsq}{\nav}-1+z\nav u(c(A\nav/\rho)^{1/3})\right]
\end{equation}  
where $\navsq=\sum_{n_s} p_{n_s} n_s^2$ is the second moment of the cluster size distribution.

One notices that the energy density worked out above only depends on
two moments of the distribution of cluster sizes $n_s$. The system
will adopt a configuration that minimizes the energy density at given
$\rho$, and we can think of this as a two-step process of first
minimizing w.r.t.\ $\navsq$ at fixed $\nav$, and then w.r.t.\ $\nav$.
The first step here can be shown to have the intuitively obvious
result that only two cluster sizes occur, namely the integers either
side of $\nav$, which we write as $\lfloor \nav\rfloor$ and $\lceil \nav \rceil$. The
relative weight of these is then fixed by $\nav$, and one finds
$\navsq = \nav^2 + \D(1-\D)$ where $\D=\nav-\lfloor \nav \rfloor$ lies
between zero and one.
\begin{figure}[h]
\includegraphics[type=pdf,ext=.pdf,read=.pdf,width=0.95\columnwidth,clip=true]{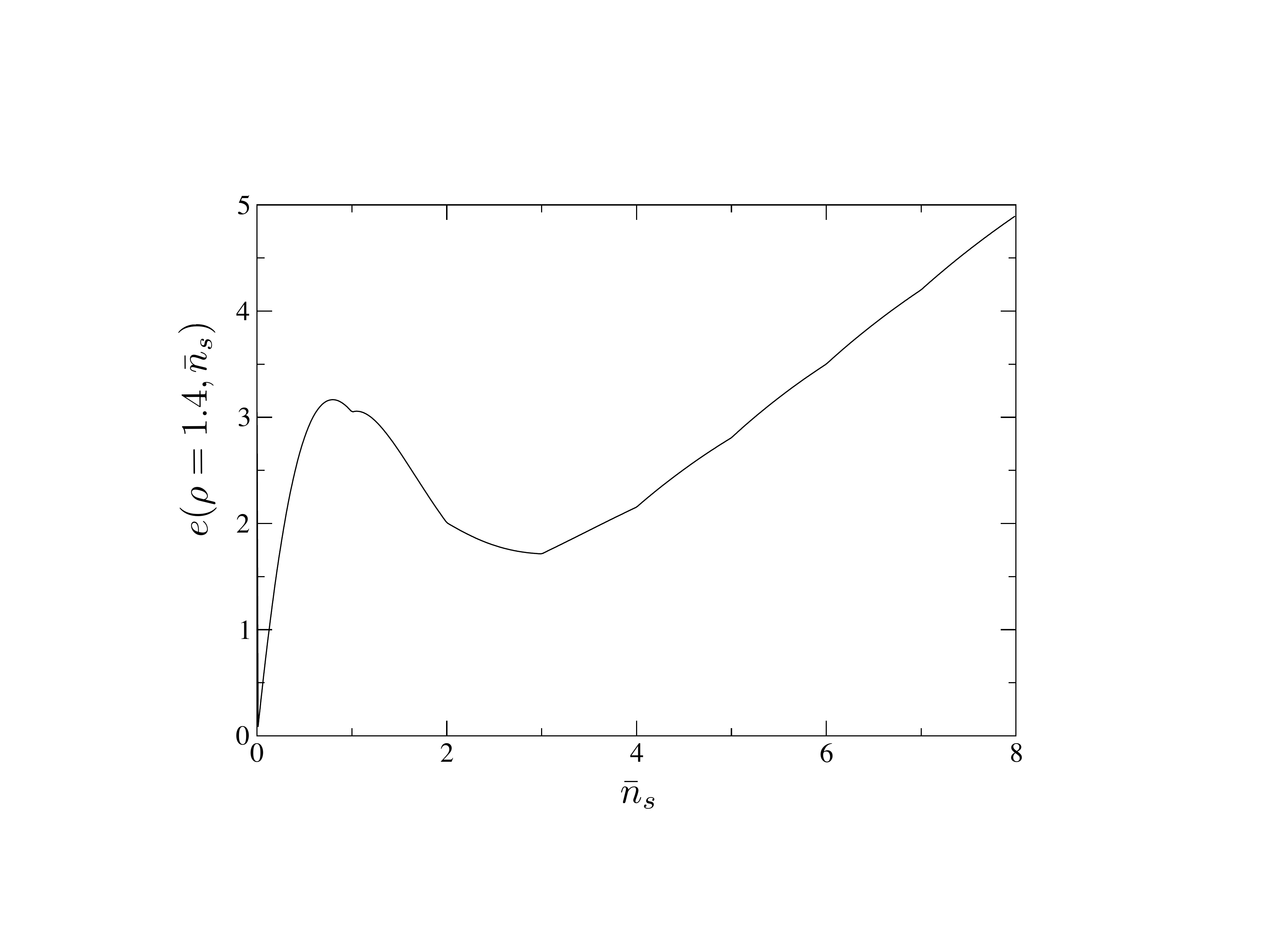}
\caption{Example plot of cell model energy density $e(\rho,\nav)$ vs average cluster size $\nav$, for potential exponent $n=4$ and density $\rho=1.4$.}
 \label{fig:enfunc}
 \end{figure}

\begin{figure}[h]
\includegraphics[type=pdf,ext=.pdf,read=.pdf,width=0.95\columnwidth,clip=true]{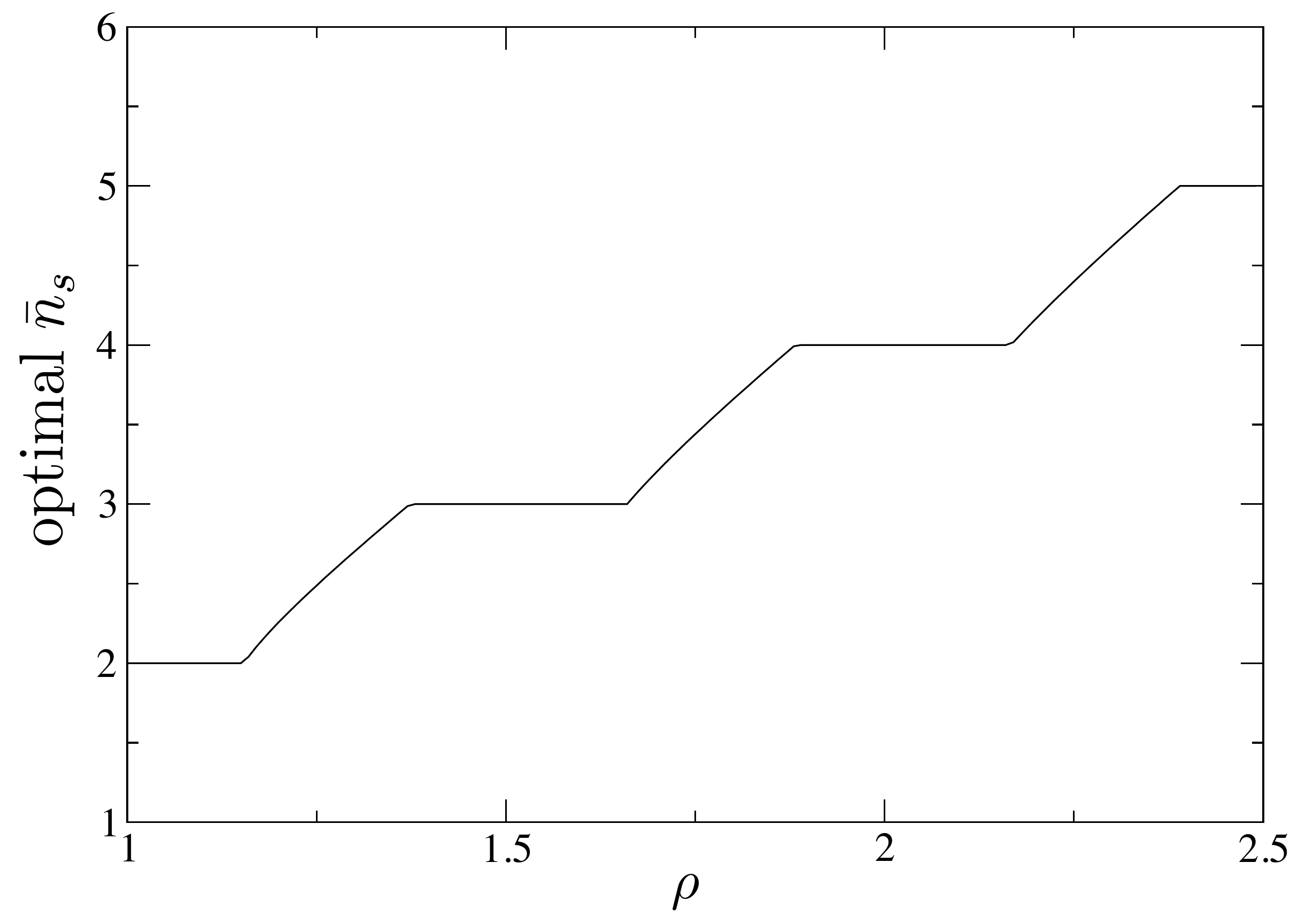}

\caption{Example plot showing for $n=4$ the value of $\nav$ as
  a function of $\rho$ that is optimal, ie.\ that minimizes the energy function (\protect\ref{e_rho_nav}).}
 \label{fig:optimal}
 \end{figure}

The cell model energy, minimized at constant $\nav$, thus becomes
\begin{equation}
e(\rho,\nav)=
\frac{\rho}{2}\left[\frac{\D(1-\D)}{\nav}+\nav-1+z\nav
  u(c(A\nav/\rho)^{1/3})\right]
\label{e_rho_nav}
\end{equation}
and the final energy density we want is $\emin(\rho)=\min_{\nav}
e(\rho,\nav)$. By way of orientation we plot $e(\rho,\nav)$ vs $\nav$
in Fig.~\ref{fig:enfunc} for exponent $n=4$ and density $\rho=1.4$. One sees  kinks at
integer values of $\nav$, which
result from the $\D(1-\D)$ term in (\ref{e_rho_nav}). As a
consequence, when we increase $\rho$ the optimal value of $\nav$ will
generally get ``stuck'' at an integer across a range of $\rho$, before
then moving smoothly to the next integer. This is shown in Fig.~\ref{fig:optimal}, where
we plot the optimal $\nav$ vs $\rho$, again
for $n=4$.

\begin{figure}[h]
\includegraphics[type=pdf,ext=.pdf,read=.pdf,width=0.95\columnwidth,clip=true]{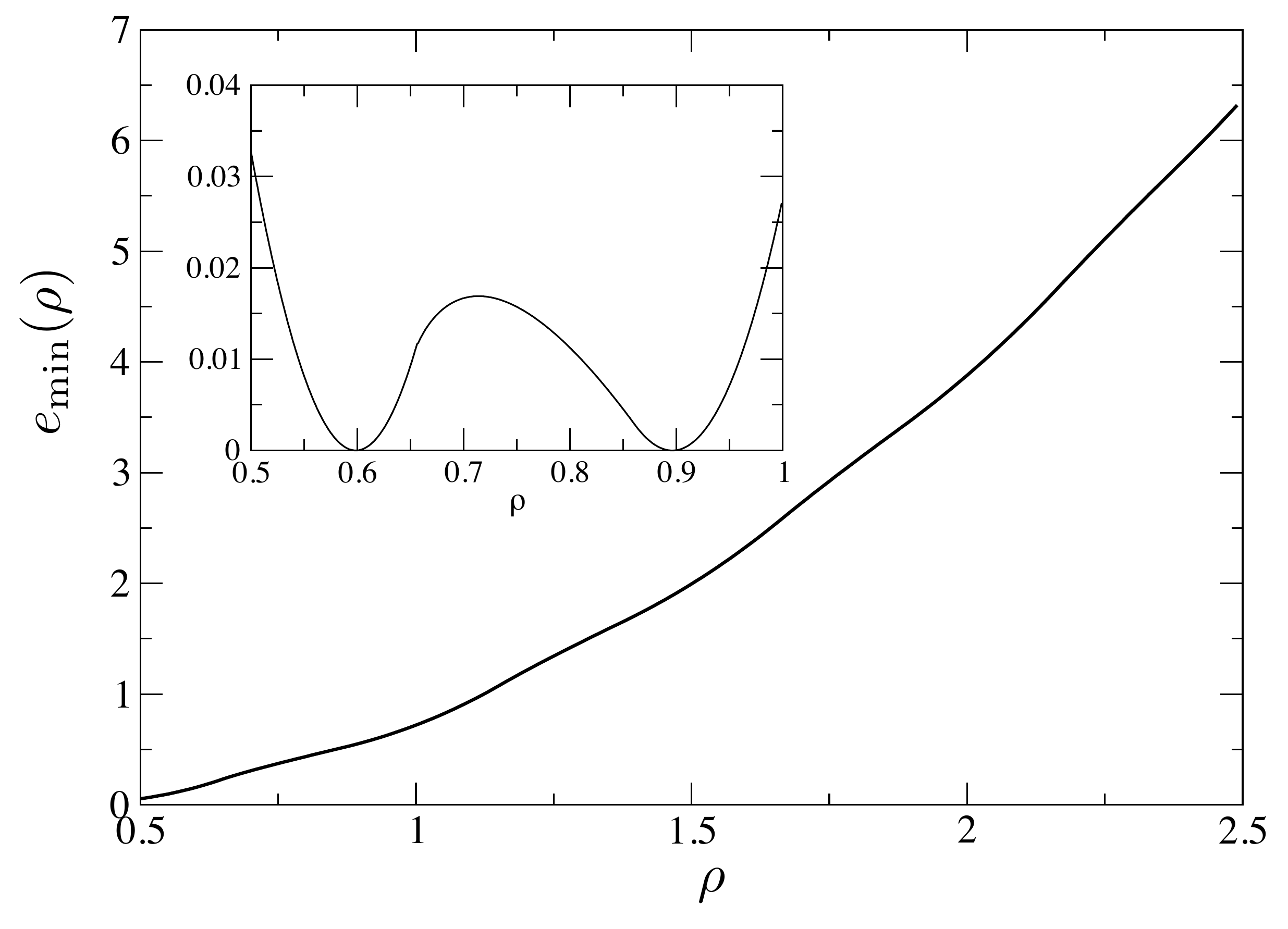}
\caption{Example plot showing for $n=4$ the form of $\emin(\rho)$ as
  described in the text. The inset plots the distance from the double
  tangent for the $\lfloor \nav\rfloor=1$ transition, which is
  explicitly $\emin(\rho)-\mu(\rho-\rho_1)-\emin(\rho_1)$ where $\mu$
  is the chemical potential at coexistence and $\rho_1$ the density of
  one of the coexisting phases.}
 \label{fig:minen}
 \end{figure}

Fig.~\ref{fig:minen} shows the resulting $\emin(\rho)$, ie.\ the minimal cell model energy density as a function of density. One sees that this consists
of a series of convex and concave regions. The existence of the
concave regions means the system will lower its energy by macroscopic
phase separation, in density regions which can be found by
constructing double tangents to $\emin(\rho)$. On general geometrical grounds the double tangents
have to touch $\emin(\rho)$ in places where the function is convex: we have checked
that these are exactly the regions where $\nav$ is an integer. The
concave regions, which are the ones where $\nav$ is not an integer,
then do not matter for the construction of the double tangents. One
thus sees a posteriori that an analysis that does not allow
substitutional disorder and assumes a fixed $n_s$ in each possible
crystal phase would have given the same result.

Allowing for substitutional disorder becomes important in the PSM
limit $n\to \infty$, however. As we explain below, in this
limit $\emin(\rho)$ approaches a function consisting of successive
straight line segments. These are already
double tangents and so the system cannot lower its energy further by
macroscopic phase separation. Therefore the equilibrium state at zero
temperature should be a substitutionally disordered crystal, in
agreement with \cite{zhangkai:2012gb}. 

To see the behaviour in the $n\to\infty$ limit, one uses the fact that
the interaction potential $u(r)$ then becomes a step function, ie.\
$=0$ for $r>1$ and $=1$ for $r<1$. As long as nearest neighbour
particles do not overlap, the $u(\cdot)$ term therefore drops out from
the PSM energy density. The resulting expression simplifies to
\begin{eqnarray}
e_\infty(\rho,\nav)&=&
\frac{\rho}{2}\left[\frac{\D(1-\D)}{\nav}+\nav-1\right]
\\
&=& \rho \left[ \lfloor \nav\rfloor -
  \frac{\lfloor\nav\rfloor(\lfloor\nav\rfloor+1)}
{2(\lfloor\nav\rfloor+\D)}\right]
\label{e_infty_rho_nav}
\end{eqnarray}
This is clearly an increasing function of $\Delta$ for each fixed
$\lfloor\nav\rfloor$, and continuous at integer values of $\nav$, so
increasing overall. Therefore the optimal value of $\nav$ is the
lowest one that is possible while maintaining the nearest neighbour
separation $d>1$ -- as is also reasonable from physical intuition --
giving $\nav=\rho/(Ac^3)$. Noting that $\lfloor\nav\rfloor+\D=\nav$,
one then sees that the first term in the square brackets in
(\ref{e_infty_rho_nav}) is linear in $\rho$, while the second one is
constant as long as $\lfloor\nav\rfloor$ remains the same. This shows
that $e_\infty(\rho)$, the minimum of $e_\infty(\rho,\nav)$ over $\nav$, is a
piecewise linear function of density. The linear segments are delimited by
integer values of $\rho/(Ac^3)$, which for fcc specifically is
$\rho/\sqrt{2}$.

\section{Results}
\label{sec:results}

We have employed the MC scheme of Sec.~\ref{sec:MC} to locate the
critical points of low-density levels of the demixing cascade for a
selection of GEM-n potentials.  Our criterion for estimating the
critical parameter was to tune the temperature and equilibrium
chemical potential until the distribution of the fluctuating number density
$p(\rho)$ closely matched the universal Ising form, which is expected
to pertain for systems with short ranged interactions and a scalar
order parameter \cite{Wilding1995,Zhang:2012ek,Wilding:2013qv}. As the
critical points can occur at very low temperature (particularly at
large $n$), relaxation times for our Monte Carlo simulations were
generally rather long. This prevented us performing a full finite-size
scaling analysis, which would have allowed us to obtain even more precise
estimates of critical point parameters. It also prevented us from
reaching the PSM limit, with $n=14$ being the steepest
potential for which we could access the critical region (see
Fig.~\ref{fig:GEM-n}).

Our cell model calculations are applicable to the zero temperature
limit in which a first order phase transition occurs, and therefore do
not provide estimates of the critical temperature. However, since the
critical temperatures of the transitions are very small, it is reasonable
to expect that the model predictions for the transition pressure and
chemical potential should be in reasonable accord with the critical
values, or at least correctly reproduce trends with respect to
variations in $n$ and the level of the cascade.

\begin{figure}[h]
\includegraphics[type=pdf,ext=.pdf,read=.pdf,width=0.95\columnwidth,clip=true]{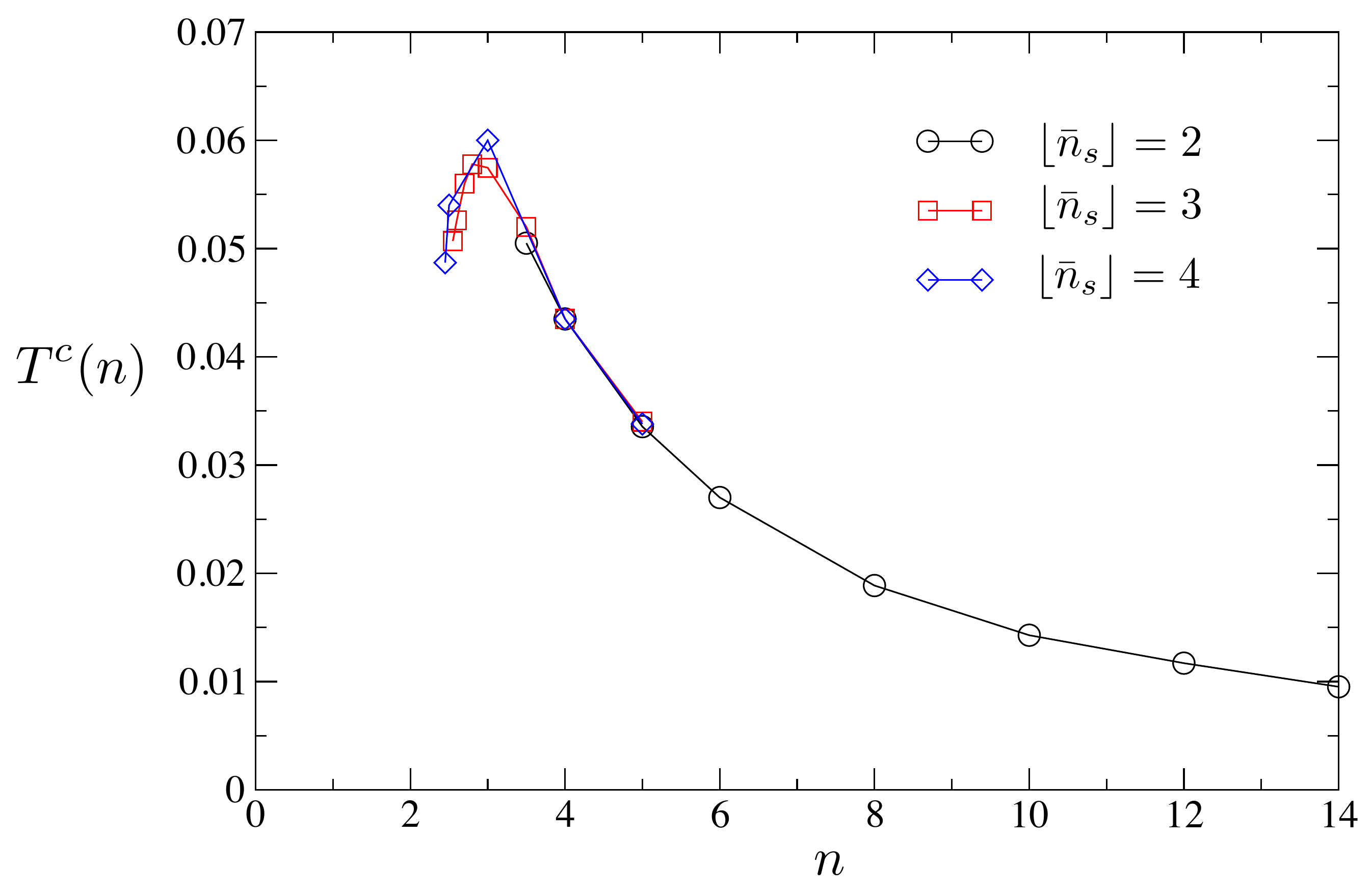}
\caption{Simulation estimates of the cascade critical temperatures for levels $\lfloor\nav\rfloor=2,3,4$ for a selection of values of the softness parameter $n$. Statistical errors are smaller than the symbol sizes.}
 \label{fig:Tc}
 \end{figure}

We consider the dependence of the critical point parameters on
the softness parameter $n$ and the cascade level, which we index by
$\lfloor \nav \rfloor$, ie.\ by the occupancy at $T=0$ of the lower
density phase of the two coexisting phases.  Estimates of the critical
temperature $T^c(n,\lfloor \nav \rfloor)$ have been made for the first
three stable levels of the cascade, corresponding to $\lfloor \nav
\rfloor=2,3,4$ and for a range of values of $n$. The results
(Fig.~\ref{fig:Tc}) show that for levels $\lfloor \nav \rfloor=3,4$
there is a maximum in $T^c$ for $n\approx 3$. No such maximum occurs
for level $\lfloor \nav \rfloor=2$, however, because on reducing $n$,
the system melts before the maximum is reached. In fact all levels of
the cascade melt as $n$ is reduced towards $n=2$. This reflects the
fact that as $n$ becomes smaller, the liquid region of the phase
diagram expands to ever greater densities, thereby engulfing
successive levels of the cascade. Such an observation is consistent
with the known phase behaviour of the GCM ($n=2$) for which 
no cluster crystals have been observed \cite{Prestipino:2005uq,Ikeda:2011qy}.

\begin{figure}[h]
\includegraphics[type=pdf,ext=.pdf,read=.pdf,width=0.95\columnwidth,clip=true]{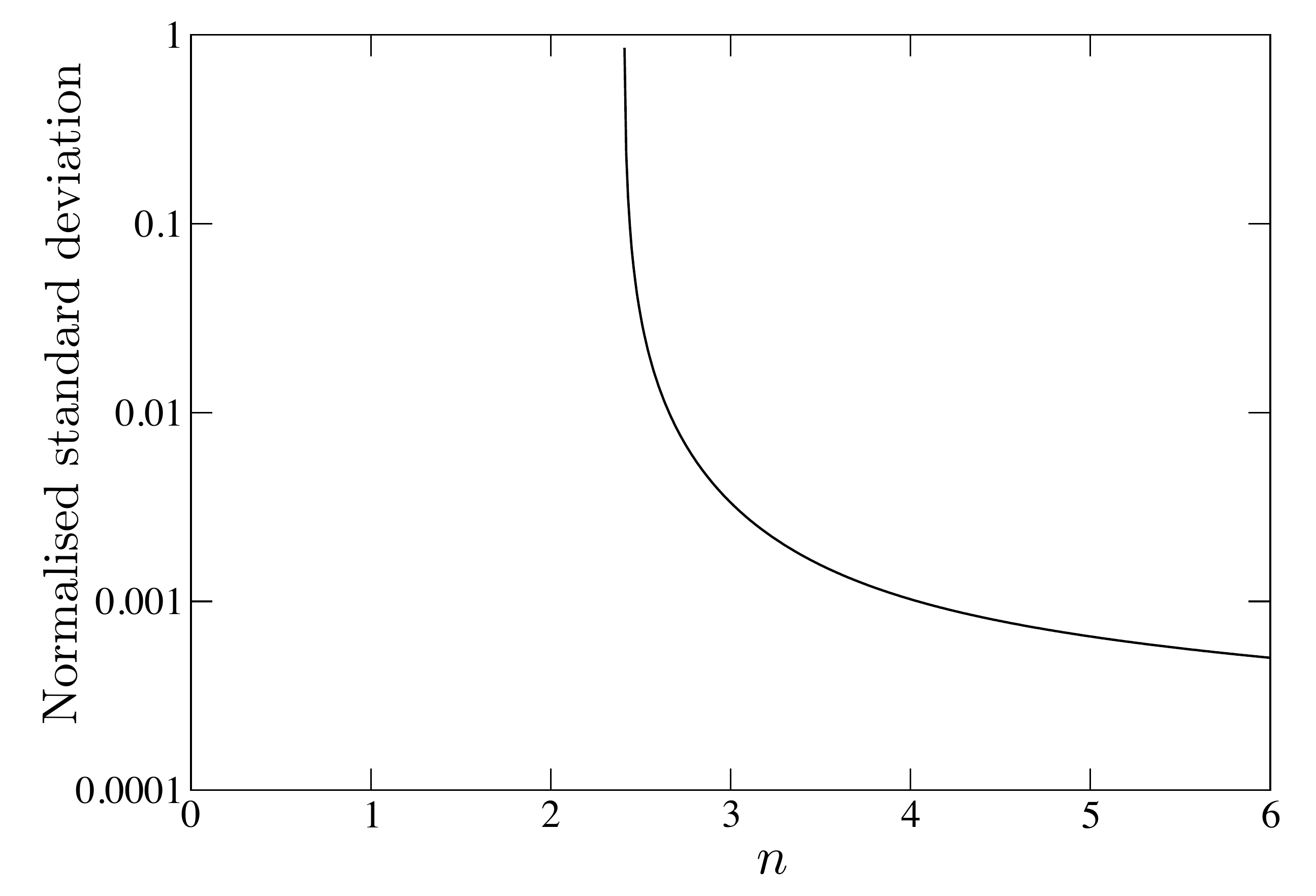}
\caption{The normalized standard deviation in the density difference order parameter at $T=0$ across
  levels $\lfloor\nav\rfloor=2,3,4$, plotted as a function of the softness parameter $n$.}
 \label{fig:densgap}
 \end{figure}

The results of Fig.~\ref{fig:Tc} exhibit the further interesting
feature that for a given $n$, the critical temperatures of each level
of the cascade are indistinguishable within uncertainty for
$n\ge4$. By contrast for $n< 4$, $T^c$ clearly increases between
levels $3$ to $4$. To help shed light on this observation, we have
used the cell model to calculate the magnitude of the density
difference (ie.\ the order parameter) for the transitions of the
cascade, at $T=0$. Since this order parameter is expected to depend on
the value of the critical temperature, it should provide an analytical
indicator as to whether $T^c$ is really independent of
$\lfloor\nav\rfloor$ for a given $n$.  Fig.~\ref{fig:densgap} shows the
standard deviation (normalised by the mean) in the value of the order parameter at $T=0$ for
levels $\lfloor \nav \rfloor=2,3,4$. One notes that this quantity is
very small across the board, in accord with the simulation findings
that variations in $T^c$ with $\lfloor\nav\rfloor$ are small. However
the variation remains non zero, and for small $n$ is considerably
stronger than for large $n$. Thus it seems likely that the apparent
independence of $T^c(n)$ on $\lfloor \nav \rfloor$ for $n\ge 4$ merely
reflects the fact that the temperature variation is smaller than can
be resolved in our simulations.

\begin{figure}[h]
\includegraphics[type=pdf,ext=.pdf,read=.pdf,width=0.95\columnwidth,clip=true]{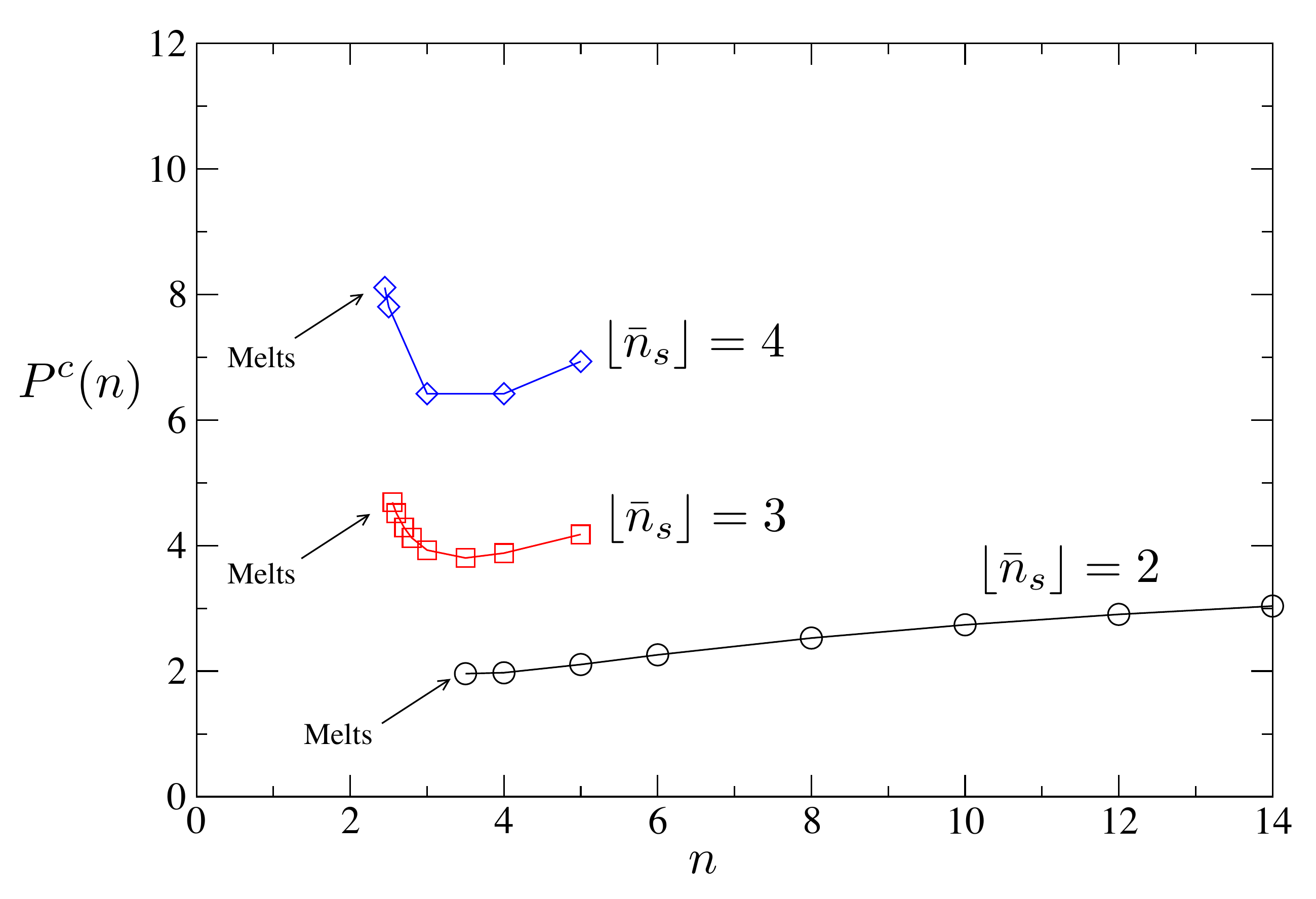}
\includegraphics[type=pdf,ext=.pdf,read=.pdf,width=0.95\columnwidth,clip=true]{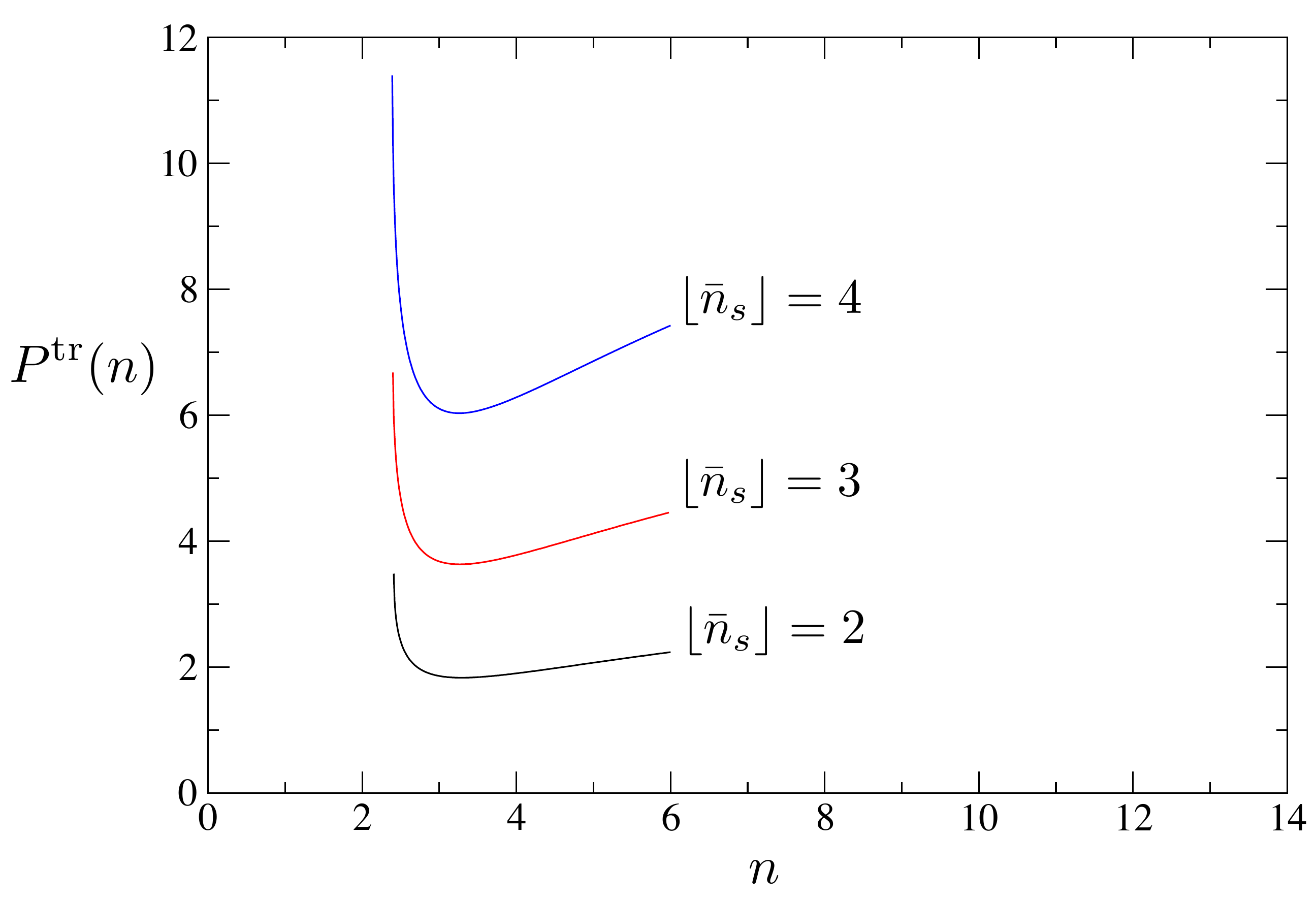}
\caption{{\bf (a)} Simulation estimates of the critical pressure
  $P^c(n)$ for the first three levels of the demixing cascade for a
  selection of values of $n$. {\bf (b)} Cell model predictions of the
  transition pressure as a function of $n$ at zero temperature for the first three levels of
  the demixing cascade.}
 \label{fig:Pc}
 \end{figure}

Figure~\ref{fig:Pc}(a) shows the simulation estimates for the critical
pressure $P^c(n)$. In contrast to the case of the critical
temperature, for any given $n$ there are large differences in the
pressure between successive levels of the cascade.  In common with the
situation for the critical temperature, the pressure varies non
monotonically in $n$, with a clear minimum close to $n=3$.  We note
that on reducing $n$ below $n=3$, the critical pressure for the
$\lfloor\nav\rfloor=3$ and $\lfloor\nav\rfloor=4$ levels starts to
increase very rapidly, before the system melts. Similar behaviour is
seen in the cell model predictions for the phase transition pressure
at $T=0$ (Figure~\ref{fig:Pc}(b)). Here the curve of the coexistence
pressure versus $n$ terminates at some value of $n$ below which no
double tangent in $\emin(\rho)$ can be found.  The value of $n$ for
which this happens lies around $n\approx 2.4$ and is only weakly
dependent on the level of the cascade. This termination presumably
reflects the instability of the cluster crystal phase.

\begin{figure}[h]
\includegraphics[type=pdf,ext=.pdf,read=.pdf,width=0.95\columnwidth,clip=true]{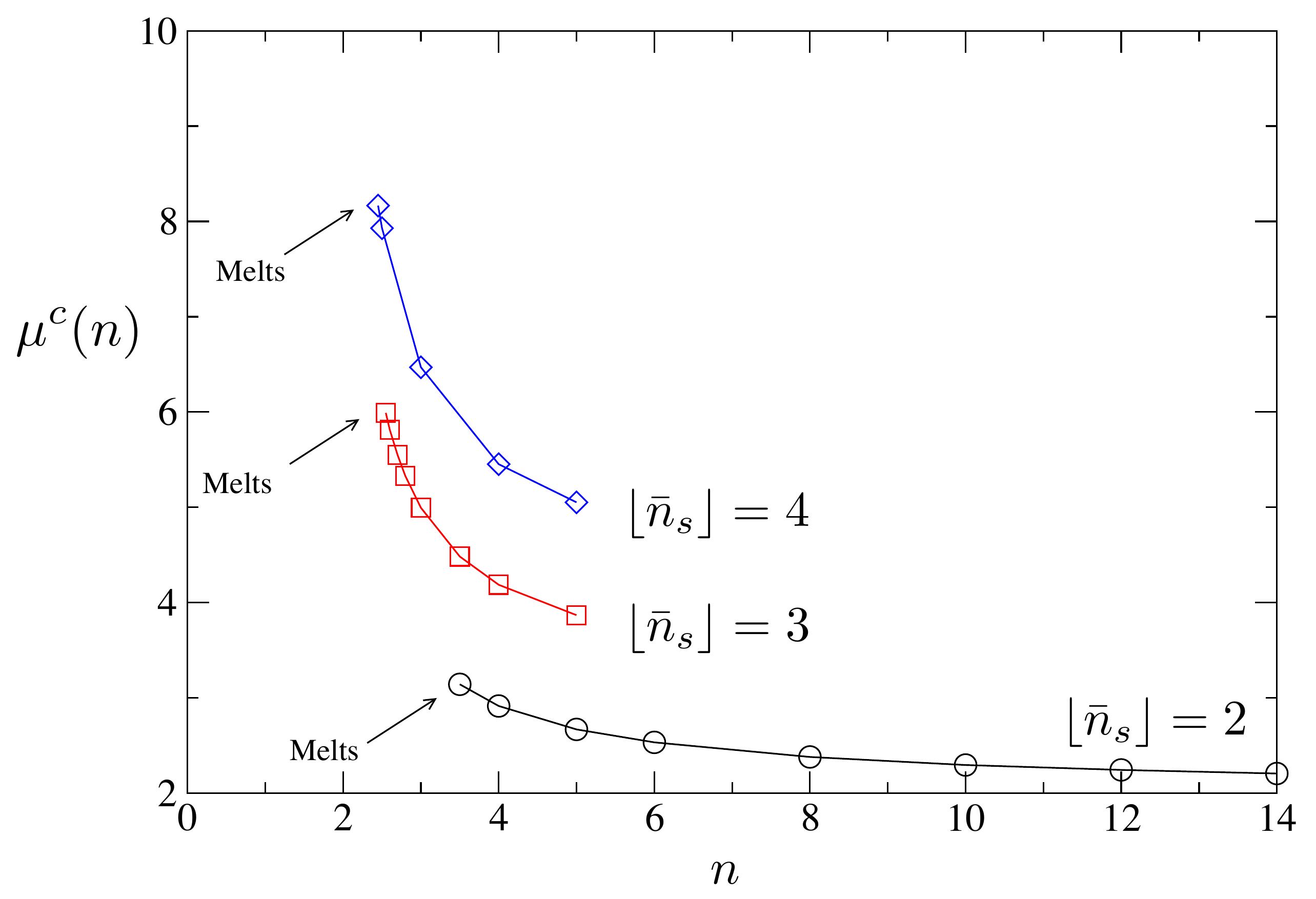}
\includegraphics[type=pdf,ext=.pdf,read=.pdf,width=0.95\columnwidth,clip=true]{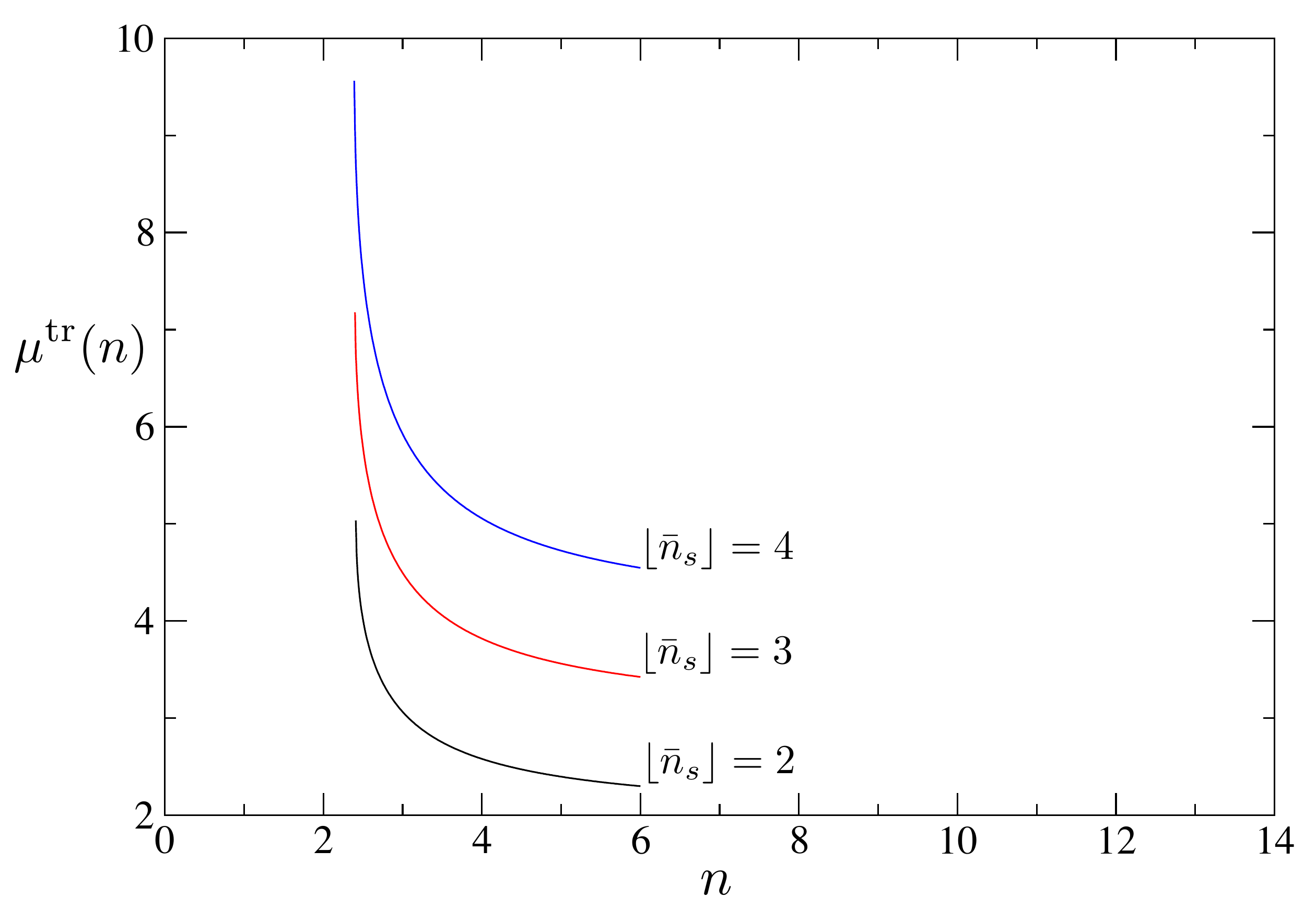}
\caption{{\bf (a)} Simulation estimates of the critical chemical
  potential $\mu^c(n)$ for the first three levels of the demixing
  cascade for a selection of values of $n$. {\bf (b)} Cell model
  predictions of the transition chemical potential as a function of
  $n$ at zero temperature for the first three levels of the demixing
  cascade.}
 \label{fig:muc}
 \end{figure}

In contrast to the scenario observed for the critical temperature and
pressure, monotonic behaviour is seen in the
critical chemical potential $\mu^c(n)$. The results (Fig.~\ref{fig:muc})
demonstrate that $\mu^c$ simply increases ever more rapidly as $n$
decreases until the system melts. Similar behaviour is observed for the transition chemical
potential at $T=0$ within the cell model, with the curves
for the transition chemical potential $\mu^{\rm tr}(n)$ terminating at $n\approx 2.4$.

\begin{figure}[h]
\includegraphics[type=pdf,ext=.pdf,read=.pdf,width=0.95\columnwidth,clip=true]{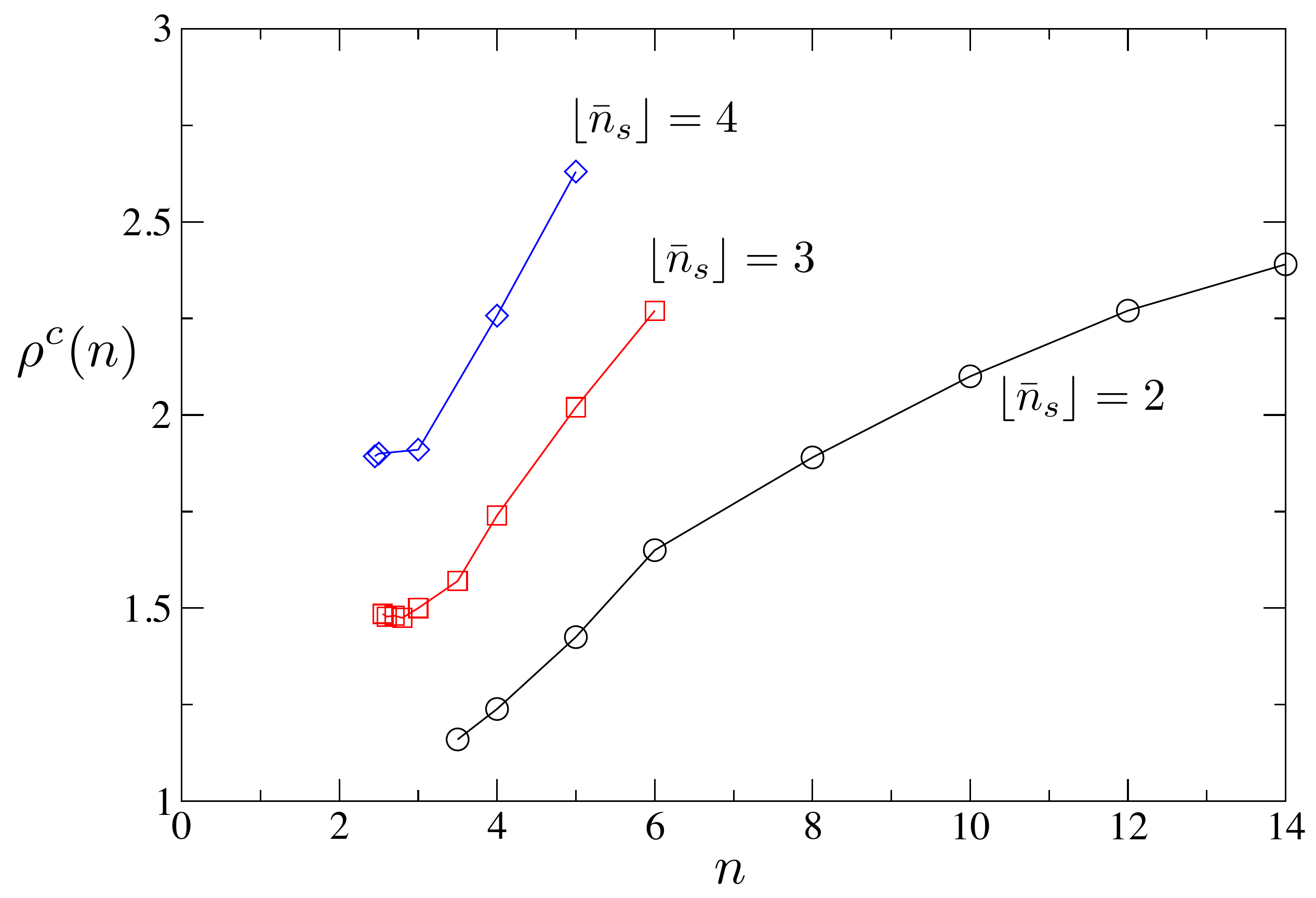}
\includegraphics[type=pdf,ext=.pdf,read=.pdf,width=0.95\columnwidth,clip=true]{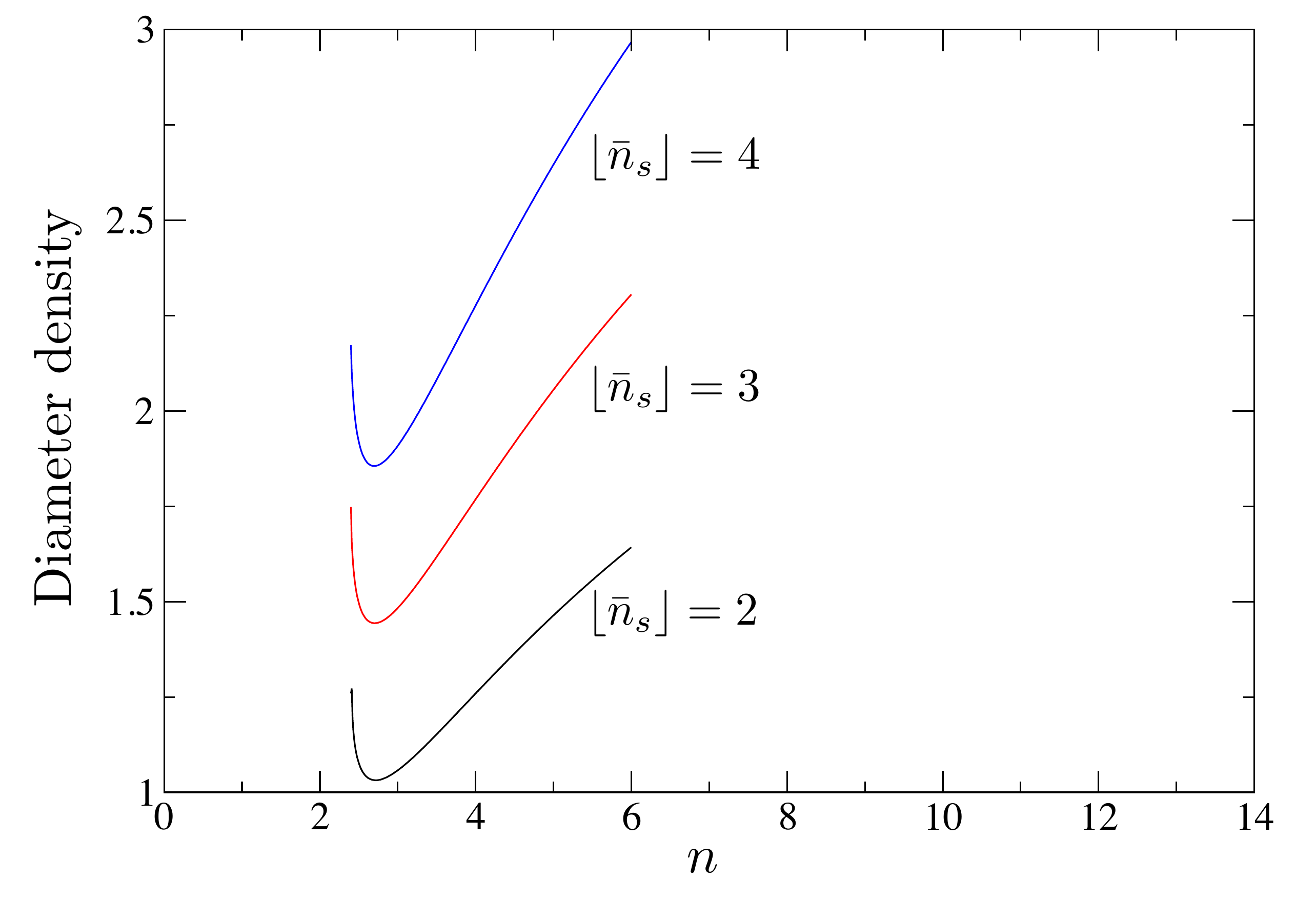}

\caption{{\bf (a)} Estimates of the critical density $\rho^c$ for
levels $\lfloor \nav \rfloor=2,3,4$ of the demixing cascade for a
selection of values of the potential softness parameter
  $n$. Statistical errors are comparable with the symbol sizes. {\bf
    (b)} Cell model predictions of the coexistence diameter density as
  a function of $n$ at $T=0$ for the first three levels of the
  demixing cascade.}
 \label{fig:rhoc}
 \end{figure}

Simulation estimates of the dependence of the critical density on $n$ and $\lfloor \nav
\rfloor$ are plotted in in Fig.~\ref{fig:rhoc}(a). Here one sees hints of the approach to a
minimum, at least for $\lfloor \nav \rfloor=3,4$, although the actual
minimum seems to be preempted by melting. Clear minima are visible, however, in
the cell model results for the coexistence diameter density at $T=0$,
Fig.~\ref{fig:rhoc}(b).

\begin{figure}[h]
\includegraphics[type=pdf,ext=.pdf,read=.pdf,width=0.95\columnwidth,clip=true]{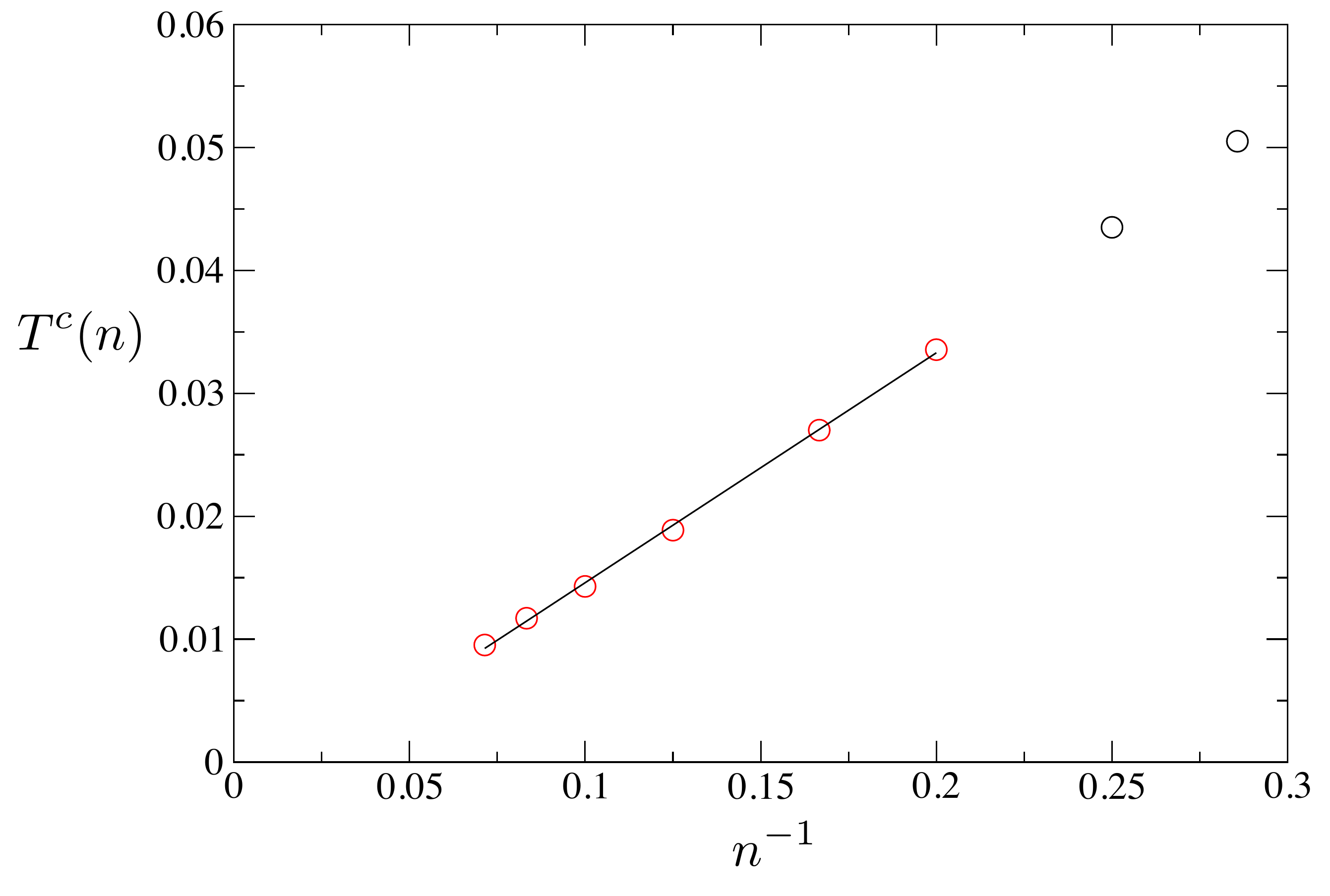}

\caption{Plot of the critical temperature of the $\lfloor \nav
\rfloor=2$ transition as a function of $1/n$. A linear fit is shown
to the data for $n>4$.}
 \label{fig:Tcscale}
 \end{figure}

Finally, we consider the behaviour of the critical temperature at
large $n$.  Fig.~\ref{fig:Tcscale} replots our simulation estimates of $T^c(n)$
for the lowest stable level of the cascade, $\lfloor \nav \rfloor=2$,
this being the level for which we were able to scan the largest range of
$n$. These results show that $T^c$ decreases rapidly with
increasing $n$. At large $n$ (ie.\ well away from the peak in
Fig~\ref{fig:Tc}) we observe scaling consistent with $T^c\sim
n^{-1}$. An extrapolation of the trend is consistent with the absence
of a demixing cascade in the PSM at zero temperature, as suggested by
the cell model studies, although since the largest $n$ we could study
was $n=14$, our data would not by themselves completely rule out a very low temperature
critical point.

\section{Summary and conclusions}

In summary, we have used tailored Monte Carlo simulation techniques
and zero temperature cell model calculations to study the behaviour of
the demixing cascade as a function of the potential softness parameter
in the GEM-$n$ models. For a given $n$, the critical temperature is
only very weakly dependent on the level of the cascade, with the
differences being greatest for $n<3$. A maximum occurs in $T^c(n)$ near
$n=3$ and non-monotonicity is also observed in the critical pressure
$P^c(n)$, but not in the critical chemical potential $\mu^c(n)$. These
latter features are corroborated by our cell model calculations for the
$T=0$ transitions.

As $n\to2^+$ (the GSM limit) the liquid region of the phase diagram
expands to higher densities. This results in the melting of successive
levels of the cascade. However, it is an interesting
open question whether the liquid always wins in the GSM limit or
whether at extremely high densities a cluster crystal can nevertheless
occur.

As $n$ is increased to large values, the critical temperatures fall
steadily to very low values, with the simulations suggesting
$T^c(n)\sim n^{-1}$.  Extrapolation of the results to the PSM
($n=\infty$) limit is consistent with the absence of a demixing
cascade in the PSM.

\acknowledgments

We thank Rob Jack, Christos Likos and Bianca Mladek for helpful discussions.

\bibliography{/Users/pysnbw/Dropbox/Papers}
\end{document}